# In-class vs. online administration of concept inventories and attitudinal assessments

Manher Jariwala[1], Jada-Simone S. White[2], Ben Van Dusen[2], and Eleanor W. Close[3]
[1]Department of Physics, Boston University, 590 Commonwealth Avenue, Boston, MA, 02215, USA
[2]Department of Science Education, California State University Chico, 101 Holt Hall, Chico, CA, 95929, USA
[3]Department of Physics, Texas State University, 601 University Drive, San Marcos, TX, 78666, USA

This study investigates differences in student responses to in-class and online administrations of the Force Concept Inventory (FCI), Conceptual Survey of Electricity and Magnetism (CSEM), and the Colorado Learning Attitudes about Science Survey (CLASS). Close to 700 physics students from 12 sections of three different courses were instructed to complete the concept inventory relevant to their course, either the FCI or CSEM, and the CLASS. Each student was randomly assigned to take one of the surveys in class and the other survey online using the LA Supported Student Outcomes (LASSO) system hosted by the Learning Assistant Alliance (LAA). We examine how testing environments and instructor practices affect participation rates and identify best practices for future use.

## I. INTRODUCTION

Concept inventories, such as the Force Concept Inventory (FCI) [1] and the Conceptual Survey of Electricity and Magnetism (CSEM) [2], are assessments designed to measure students' knowledge of a concept that is core to a discipline. Using concept inventories as pre- and post-test, at the beginning and end of a course, has become a common method for assessing student learning during a course. Having concept inventories as tools that are research-based, i.e., have validation arguments and nationally normed outcome data associated with them, has been a significant driver of both physics education research and course transformation [3].

While the use of research-based concept inventories and attitudinal assessments has spread in the physics community over the two decades since the creation of the FCI, there are still many physics faculty who do not use them in their classes. There are a number of reasons that faculty choose not to use concept inventories, including not knowing what assessments exist, not wanting to use class time for administering them, and having difficulty in analyzing student results. Several online resources, such as PhysPort [4] and its DataExplorer tool [5], have been created to alleviate difficulties of using concept inventories.

In an effort to increase the use of concept inventories and PER advancement, the Learning Assistant Alliance, an international network of LA-using institutions [6], created the Learning Assistant Supported Student Outcomes (LASSO) tool [7]. LASSO is a free online tool for administering, scoring, analyzing, and tracking students' concept inventory scores [8]. To use LASSO, faculty answer a few questions about their course, select any concept inventories they wish to administer, and upload a class list. Faculty can then launch and close the pretest window at their discretion. Each student receives an email with instructions and a unique link to the online concept inventory (which has been assigned a generic name to safeguard the integrity of the assessment). Faculty can easily track student participation and send out reminder emails with the click of a button. At the end of the semester, the post-test assessment is done identically to the pretest. At the end of the class, faculty can download a spreadsheet with their students' scored answers as well as a PDF report analyzing their students' learning. In addition to providing assistance interpreting assessment outcomes, the online platform for concept inventories allows faculty to avoid using class time and, instead, to offer students homework or other credit as incentive.

Online tools can make it easier for faculty to administer concept inventories, however, there are potential issues with transitioning assessments from being administered with paper and pencil in class to being done online at home. Common concerns about this transition include how it will impact student participation rates, test validity, and test security. The research presented in this paper is part of an ongoing project designed to investigate each of these concerns. The first step in this process, and the focus of this paper, is investigating student participation rates. Once a set of best practices for increasing student participation on online concept inventories is established, student performance will be examined.

## II. RESEARCH QUESTIONS

This mixed-methods investigation examines student use of the LASSO system and summarizes faculty reports and interviews to answer the following questions: (1) How does the use of online concept inventories impact student participation rates, if at all? (2) How do faculty practices impact student use of online concept inventories, if at all?

## III. METHODS

Data for this investigation were collected from three introductory physics courses with five instructors and almost 700 students (693 to start, 659 at end, 5% attrition) at a large public university. An experimental design was used that assigned students into one of two testing conditions. The first testing condition asked students to complete a Concept Inventory (either the FCI [1] or the CSEM [2]) in class and the Colorado Learning Attitudes about Science Survey (CLASS) [9] online using the LASSO tool (Table I). The second testing condition reversed which instruments were completed in class and online. A stratified random sampling technique was used that divided students from each section into gender, racial, and honors student groups. Within each of these groups, students were randomly assigned testing conditions to ensure that each condition would have proportional representation from each group.

TABLE I. Course descriptions, assessments given, and number of students at start of semester, by instructor.

| Instr. | Course Description | Assessments Given | N to start |
|---|---|---|---|
| 1 | Algebra-based mechanics | FCI, CLASS | 320 |
| 2 | Calculus-based E&M | CSEM, CLASS | 89 |
| 3 | Calculus-based E&M | CSEM, CLASS | 50 |
| 4 | Calculus-based mechanics | FCI, CLASS | 46 |
| 5 | Calculus-based mechanics | FCI, CLASS | 188 |

Instructors were informed of which students were in each testing condition but were not provided a specific script to follow in the administration of their assessments. At the conclusion of the semester, instructors were interviewed to determine their methods for administering the pre- and post-tests in both paper and online mode. For each section within a course, participation rates of both categories (paper and online) were summarized for each instrument (FCI, CSEM, or CLASS). All students took (both) one assessment online (CLASS or CSEM/FCI) and one assessment in class (CSEM/FCI or CLASS). Because there was very similar participation between the instruments administered online within a given course section, we binned both assessments into a single "online" category. For the paper assessments, half of the class received the CLASS (attitudinal survey) and half of the class received the FCI / CSEM concept inventory (depending on course type) at the same time. The testing environment was constant and we are confident in binning these data into a single "paper" category. Given this experimental design, we found it important to conduct faculty interviews to determine similarities or differences in how students were motivated to participate in the online assessments.

Variations in participation rates were measured across testing conditions and instructors. Statistical significances between conditions were measured using chi-square tests with Bonferroni corrections (when applicable).

## IV. RESULTS

Participation rates are reported as the percentage of students who completed the paper and online assessments, combined across all class sections for each instructor. Overall, as shown below in Figure 1, participation rates in class were nearly twice as high as participation rates online ($p < 0.001$). In both formats, there was a drop in participation in the post-tests as compared to the pre-tests.

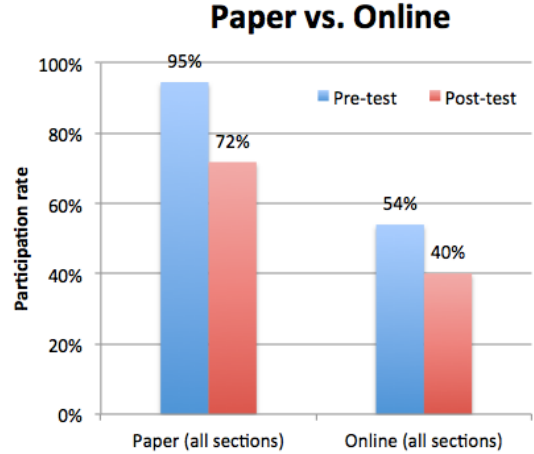

FIG 1. Pre- and post-test participation rates for assessments administered on paper and online, across all sections.

Figure 2 shows the pre- and post-test participation rates just for the paper assessments, but separated by instructor. Participation was uniformly high during pre-tests administered in class (87-97%). However, despite in-class administration, post-test participation varied widely across instructors (55-95%), suggesting that differences in the details of administering assessments among instructors significantly impacted student participation.

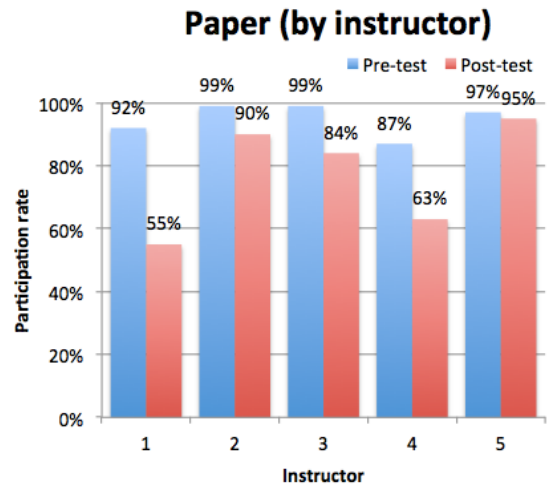

FIG. 2. Pre- and post-test participation rates for all assessments administered on paper, by instructor.

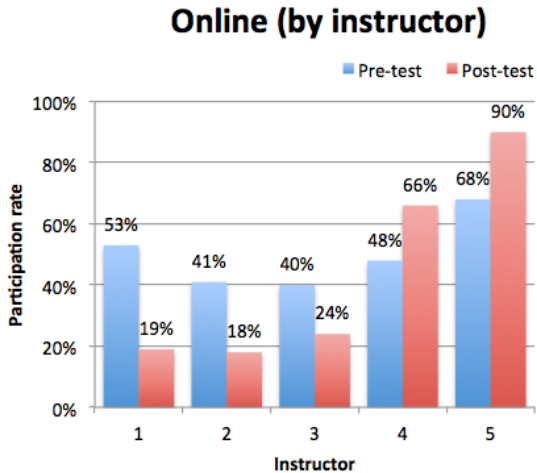

FIG. 3. Pre- and post-test participation rates for all assessments administered online, by instructor.

Figure 3 shows the pre- and post-test participation rates for the online assessments, also separated by instructor. Online participation in pre-tests was uniformly mediocre (40-53%), with one exception (instructor 5: 68%; Fig. 3). Online post-test participation was substantially lower than the pre-test participation for three instructors (18-24%), while it increased substantially for the other two (66-90%), with consistently higher online participation for instructor 5 across pre- and post-tests, relative to the others (Fig. 2).

While the participation rates were generally lower for online conditions, participation rates within single instructor's courses showed different patterns. Instructors 1-3 had statistically significant lower participation rates on the online pre and posttests ($p < 0.001$) as compared to paper. Instructors 4 and 5 had statistically significant lower online pre-test participation rates ($p < 0.001$) as compared to paper, but did not have statistically significant differences in post-test participation rates between their paper and online conditions.

## V. DISCUSSION

In order to understand the differences in participation rates, we interviewed each of the instructors to determine the details of how they administered their assessments, starting with the assessments given on paper, in-class.

### A. Paper practices

Table II summarizes these findings for the assessments given on paper, indicating (i) whether students who missed the in-class assessment were allowed to make it up at a later date and (ii) whether extra credit was offered as a way to encourage students to complete the assessment.

As shown previously in Fig. 2, all five instructors had typically high participation rates for their paper pre-tests, given in class on or near the first day of the semester. This can largely be explained by attendance, which during the first week of the semester is typically close to 100%. Three of the instructors (labeled 2, 3, and 5) also offered makeups for the paper pre-test for those students who missed it; it should be noted that instructor 5 administered the paper pre-tests for the students of instructors 2 and 3, while also offering the makeups for students in all three classes. Correspondingly, these three classes show even higher participation rates for the paper pre-tests (99%, 99%, 97%) than the other two classes that did not offer makeups (92%, 87%). Extra credit was not offered for any of the paper pre-tests, as also shown in Table II; "No*" indicates that the instructor 1 later added extra credit to the scores of students who completed any assessments, but did not advertise this in advance to any students as a means of encouraging them to participate.

TABLE II. Paper pre- and post-test practices, by instructor, indicating whether makeups were allowed and if extra credit was given for completing the assessment.

| Instr. | Pre Makeups | Pre Ex. Credit | Post Makeups | Post Ex. Credit |
|---|---|---|---|---|
| 1 | No | No * | No | No * |
| 2 | Yes | No | Yes | No |
| 3 | Yes | No | Yes | No |
| 4 | No | No | **Yes** | **Yes** |
| 5 | Yes | No | Yes | **Yes** |

For the paper post-tests, Table II indicates in boldface the practices that are different than for the pre-test. Instructor 4 now offered makeups for their paper post-test, joining 2, 3, and 5. Another difference was that instructors 4 and 5 now offered and advertised extra credit for completing the paper post-test. The three classes which had higher paper pre-test participation rates had relatively high rates for the paper post-tests (90%, 84%, 95%), especially compared to the other two classes (55%, 63%). Interestingly, comparing instructors 4 and 5, similar "best practices" of offering make-ups as well as adding on extra credit resulted in very different participation rates (95% and 63%) for the paper post-test in the two calculus-based mechanics classes. This may be due to differences in the timing with which the two instructors advertised the available extra credit: instructor 5 alerted students a week before the post-test administration, and had high attendance on the day of the post-test, while instructor 4 did not give advance notice and had lower attendance.

### B. Online practices

Table III summarizes the instructor practices regarding the administration of the online pre- and post-test assessments, indicating (i) whether one or multiple email reminders were sent out to students and again (ii) whether extra credit was offered.

For the online pre-tests, as before with the paper pre-tests, none of the instructors offered any extra credit. Instructors 2 and 5 did send out multiple email reminders to students, while the other three instructors only sent out one email each (through the LASSO system). Our results for instructors 2 and 5 do not indicate that sending out multiple email reminders necessarily led to higher completion rates, perhaps due to student unfamiliarity with the online assessment system at the start of the semester.

TABLE III. Online pre- and post-test practices, by instructor, indicating how many email reminders were sent and if extra credit was given for participating.

| Instr. | Pre Emails | Pre Ex. Credit | Post Emails | Post Ex. Credit |
|---|---|---|---|---|
| 1 | One | No * | **No** | No * |
| 2 | Multiple | No | Multiple | No |
| 3 | One | No | One | No |
| 4 | One | No | **Multiple** | Yes |
| 5 | Multiple | No | Multiple | Yes |

For the online post-tests, differences in instructor practices seemed to have marked effects. For instructor 1, sending no emails at all and not offering extra credit resulted in an extremely low participation rate of 19%. Instructors 2 and 3 did not fare much better at 18% and 24%, respectively, despite sending out email reminders but not offering extra credit. Instructors 4 and 5, however, achieved the highest online post-test participation rates by utilizing multiple email reminders and extra credit. Strikingly, instructor 5 achieved an extremely high online post-test rate (90%) nearly equal to both the paper pre- and post-test rates for this class. Similarly, even though the online post-test rate for instructor 4 is lower at 66%, it is equivalent to the paper post-test rate for this class, indicating no difference in response rate with regards to the assessment delivery.

## VI. CONCLUSIONS

Our study confirms that, without being thoughtful about its implementation, online administration of research-based assessments generally suffers from lower participation rates than using paper assessments in class. Two of the instructors saw improved participation rates in the administration of the post-test and offer insight on how to improve participations rates more broadly. Instructor practices such as offering makeups, sending out multiple email reminders, and offering extra credit for completion can counteract the drop-off in participation rates from pre-test to post-test, as well the discrepancy between paper and online administration of assessments. With these changes, one instructor had an online post-test participation rate that exceeded that of the paper post-test participation rate of every other instructor. Given the growing focus on centralized, online administration of research-based assessments [10], our study offers hope that, when scaffolded properly, online assessment participation rates can rival those of in-class assessments.

Using best practices to increase online participation facilitates data collection for large-scale studies of student learning outcomes, using tools like LASSO. Future work will extend this research to include analysis of variation in student performance across in class and online assessments.

### ACKNOWLEDGEMENTS

This work was funded in part by NSF-DUE #1525354 (MJ), NSF-DUE #1525338 (JSW, BVD), and NSF #0808790 and #1431578 (EWC). This paper is contribution No. LAA-036 of the International Learning Assistant Alliance.